\documentclass[letterpaper,twocolumn,prl,aps,superscriptaddress,floatfix]{revtex4}

\usepackage[latin9]{inputenc}
\usepackage{amsmath}
\usepackage{amssymb}
\usepackage{graphicx}
\usepackage{esint}
\usepackage{times}

\usepackage[unicode=true,
 bookmarks=true,bookmarksnumbered=false,bookmarksopen=false,
 breaklinks=false,pdfborder={0 0 1},backref=false,colorlinks=true]
 {hyperref}
\hypersetup{
 linkcolor=magenta, urlcolor=blue, citecolor=blue, pdfstartview={FitH}, hyperfootnotes=false, unicode=true}

\makeatletter

\pdfpageheight\paperheight
\pdfpagewidth\paperwidth

\@ifundefined{textcolor}{}
{%
 \definecolor{BLACK}{gray}{0}
 \definecolor{WHITE}{gray}{1}
 \definecolor{RED}{rgb}{1,0,0}
 \definecolor{GREEN}{rgb}{0,1,0}
 \definecolor{BLUE}{rgb}{0,0,1}
 \definecolor{CYAN}{cmyk}{1,0,0,0}
 \definecolor{MAGENTA}{cmyk}{0,1,0,0}
 \definecolor{YELLOW}{cmyk}{0,0,1,0}
}

\usepackage{xcolor}\usepackage{soul}

\newcommand{\ket}[1]{\ensuremath{\left|#1\right\rangle}}
\definecolor{blue}{rgb}{0,0,1}
\definecolor{red}{rgb}{1,0,0}
\definecolor{green}{rgb}{0,1,0}

\makeatother

\begin{document}

\title{Observation of topological magnon insulator states in a superconducting circuit}

\author{W. Cai}
\thanks{These two authors contributed equally to this work.}
\affiliation{Center for Quantum Information, Institute for Interdisciplinary Information Sciences, Tsinghua University, Beijing 100084, China}

\author{J. Han}
\thanks{These two authors contributed equally to this work.}
\affiliation{Center for Quantum Information, Institute for Interdisciplinary Information Sciences, Tsinghua University, Beijing 100084, China}

\author{Feng Mei} \email{meifeng@sxu.edu.cn}
\affiliation{State Key Laboratory of Quantum Optics and Quantum Optics Devices,
    Institute of Laser Spectroscopy, Shanxi University, Taiyuan, Shanxi 030006, China}
\affiliation{Collaborative Innovation Center of Extreme Optics,
Shanxi University, Taiyuan, Shanxi 030006, China}

\author{Y. Xu}
\affiliation{Center for Quantum Information, Institute for Interdisciplinary Information Sciences, Tsinghua University, Beijing 100084, China}

\author{Y. Ma}
\affiliation{Center for Quantum Information, Institute for Interdisciplinary Information Sciences, Tsinghua University, Beijing 100084, China}

\author{X. Li}
\affiliation{Center for Quantum Information, Institute for Interdisciplinary Information Sciences, Tsinghua University, Beijing 100084, China}

\author{H. Wang}
\affiliation{Center for Quantum Information, Institute for Interdisciplinary Information Sciences, Tsinghua University, Beijing 100084, China}

\author{Y. P. Song}
\affiliation{Center for Quantum Information, Institute for Interdisciplinary Information Sciences, Tsinghua University, Beijing 100084, China}

\author{Zheng-Yuan Xue} \affiliation{Guangdong Provincial Key Laboratory of Quantum Engineering
and Quantum Materials, and School of Physics and Telecommunication Engineering,
South China Normal University, Guangzhou 510006, China}

\author{Zhang-qi Yin}
\affiliation{Center for Quantum Information, Institute for Interdisciplinary Information Sciences, Tsinghua University, Beijing 100084, China}

\author{Suotang Jia}
\affiliation{State Key Laboratory of Quantum Optics and Quantum Optics Devices,
    Institute of Laser Spectroscopy, Shanxi University, Taiyuan, Shanxi 030006, China}
\affiliation{Collaborative Innovation Center of Extreme Optics,
Shanxi University, Taiyuan, Shanxi 030006, China}

\author{Luyan Sun}\email{luyansun@tsinghua.edu.cn}
\affiliation{Center for Quantum Information, Institute for Interdisciplinary Information Sciences, Tsinghua University, Beijing 100084, China}

\begin{abstract}
Searching topological states in artificial systems has recently become a rapidly growing field of research. Meanwhile, significant experimental progresses on observing topological phenomena have been made in superconducting circuits. However, topological insulator states have not yet been reported in this system. Here, for the first time, we experimentally realize a tunable dimerized spin chain model and observe the topological magnon insulator states in a superconducting qubit chain. Via parametric modulations of the qubit frequencies, we show that the qubit chain can be flexibly tuned into topologically trivial or nontrivial magnon insulator states. Based on monitoring the quantum dynamics of a single-qubit excitation in the chain, we not only measure the topological winding numbers, but also observe the topological magnon edge and defect states. Our experiment exhibits the great potential of tunable superconducting qubit chain as a versatile platform for exploring non-interacting and interacting symmetry-protected topological states.

\end{abstract}

\maketitle


Topological insulators are new states of matter beyond Landau symmetry breaking theory, signified by topological invariants and topological edge states, and now lie at the forefront of condensed matter physics~\cite{Hasan2010,Qi2011}. The concept of topological insulators recently has been intensively studied and also been expanded to artificial systems, including ultracold atomic~\cite{Goldman2016,ZhangZhu2018,Cooper2019}, photonic~\cite{Lu2014,Ozawa2018} and mechanical~\cite{Huber2018,Ma2019} systems. Nevertheless, as a result of the difficulties in extracting the Berry curvature in photonic and mechanical systems and in engineering edges in optical lattices, it is still challenging to experimentally observe both the topological invariants and the topological edge states in a separate artificial topological system.

Superconducting circuits now have become one of the leading quantum platforms for implementing scalable quantum computation~\cite{You2011,Devoret2013,Neill2018} and large-scale quantum simulation~\cite{Buluta2009,Houck2012,Georgescu2014}. In particular, topological effects recently have also been experimentally studied in superconducting circuits. Specifically, topological concepts have been investigated in the parameter space of superconducting qubits~\cite{Schroer2014,Roushan2014,WangSCPMA2018,Tan2018} and the phase space of superconducting resonators~\cite{Ramasesh2017,Flurin2017}; Synthetic gauge fields~\cite{Roushan2017a,wangSR2015,ypwangnpjqi2016,Owens2018} and Hofstadter butterfly~\cite{Roushan2017b} have been realized in a superconducting qubit chain; Topological phenomena have also been observed in a network of superconducting flux qubits~\cite{King2018}. However, due to the challenges in engineering a topological chain with tunable qubit couplings and the lack of methods in detecting the topology of the qubit chain system, topological insulator states still have not been experimentally observed before in the superconducting system.

In this paper, we experimentally demonstrate the first observation of topological insulator states in a tunable superconducting transmon qubit chain, which exhibits both the nontrivial topological invariants and topological edge states. Our experiment is based on realizing a dimerized spin chain model, which supports topologically trivial or nontrivial magnon insulator states dependent on the qubit coupling configurations. We demonstrate that such configurations can be flexibly tuned via parametrical modulations of the qubit frequencies in situ~\cite{Zhou2009,Strand2013,Wu2018,Reagor2018,Li2018}. Through exciting one of the qubits in the chain and then monitoring its quantum dynamics, we further show that the topological winding numbers can be directly measured. By tuning the qubit chain with odd and even number of qubits, the localization and hybridization of topological edge states are observed respectively. Via locally tuning the qubit couplings, we also exhibit that a topological defect can be easily created and probed.


Distinct from previous systems studying the topological states of non-interacting bosons in a lattice~\cite{Goldman2016,ZhangZhu2018,Cooper2019,Lu2014,Ozawa2018,Huber2018,Ma2019}, the superconducting system allows the study of the topological states of magnons (qubit excitations) in an interacting spin (qubit) chain, where magnons are bosonic quasiparticle excitations around the ground state of the spin chain ~\cite{FukuharaNature2013,Fukuhara2013} and are interacting hard-core bosons. Although our experiment investigates the single-excitation case and observes the resulted non-interacting symmetry-protected topological magnon insulator states, it paves the way for further studying bosonic interacting symmetry-protected topological states~\cite{Chen2011Symmetry,Chen1604} when introducing multiple excitations into a longer qubit chain. Realizing interacting symmetry-protected topological states currently is still a great challenge~\cite{Sylvain2018} and can not be achieved in previously reported non-interacting topological systems~\cite{Goldman2016,ZhangZhu2018,Cooper2019,Lu2014,Ozawa2018,Huber2018,Ma2019}. Our experiment represents the first step towards realizing such states with a qubit chain system.



\begin{figure}[tb]
\includegraphics{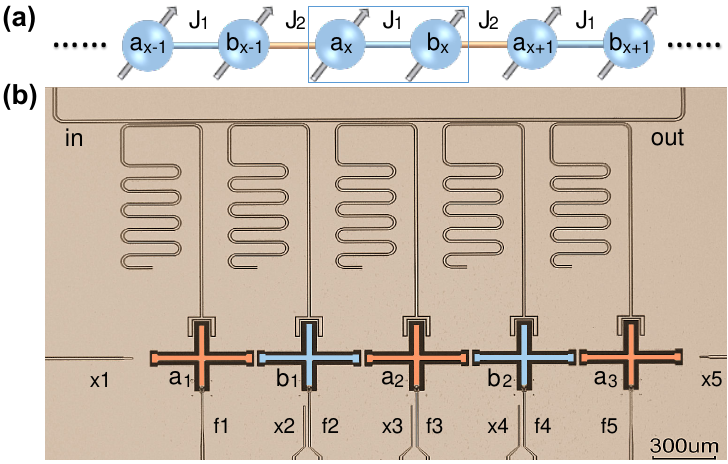}
\caption{\textbf{A dimerized qubit chain.} \textbf{a} The schematic setup of a qubit chain, where $x$ denotes the unit cell index. Each unit cell has two qubits $a$ and $b$. The intra and inter unit cell couplings are $J_1$ and $J_2$, respectively. \textbf{b} Five cross-shaped transmon qubits (Xmons, $a_1, b_1, a_2, b_2$, and $a_3$) arranged in a linear array. Each qubit is coupled to a separate $\lambda/4$ resonator for simultaneous and individual readout, and has independent $XY$ and $Z$ controls (labelled as $``x"$ and $``f"$ respectively).}
\label{fig:Model}
\end{figure}

 The experimental model is based on a dimerized spin chain which describes a one-dimensional spin lattice with two spins per unit cell and different intracell and intercell hopping amplitudes, as shown in Fig.~\ref{fig:Model}\textbf{a}. We implement such a model in a superconducting qubit chain~\cite{Mei2018}, where each unit cell contains two qubits labelled by $a$ and $b$. The resulted qubit chain can be described by the dimerized spin chain Hamiltonian after rotating wave approximation
\begin{equation}
\hat{H}=\sum_{x=1}^{N}(J_{1}\hat{\sigma}_{a_x}^{+}\hat{\sigma}^{-}_{b_x}+J_{2}\hat{\sigma}_{b_x}^{+}\hat{\sigma}^{-}_{a_{x+1}}+\text{H.c.),}
\label{Eq:DSC}
\end{equation}
where $x$ is the unit cell index, $N$ is the number of the unit cells, $J_{1}$ and $J_2$ are the intracell and intercell qubit couplings, respectively, and $\sigma^{+}_{a_x}$ ($\sigma^{-}_{a_x}$) is the raising (lowering) operator associated with qubit $a_x$. The single-qubit excitation in this spin chain is called a magnon in condensed matter physics~\cite{FukuharaNature2013,Fukuhara2013}. In the single-qubit excitation case, its topology is same as of the Su-Schrieffer-Heeger model~\cite{SSH1979,Atala2013,Meier2016,St-Jean2017}, which has two distinct topological insulator states characterized by topological winding numbers. However, the difference from the Su-Schrieffer-Heeger model is that the system studied in our experiment is an interacting spin chain and can be further used to realize interacting symmetry-protected topological states. Moreover, magnons are bosonic quasiparticle excitations around the ground state of the spin chain, therefore the topological states observed here are associated with excited states instead of ground states. When the qubit couplings are tuned into $J_1<J_2$ ($J_1>J_2$), the topological winding number $\nu=1$ ($\nu=0$) and the system supports a topologically nontrivial (trivial) magnon insulator state (see Supplementary Materials).

\begin{figure*}[tb]
\includegraphics{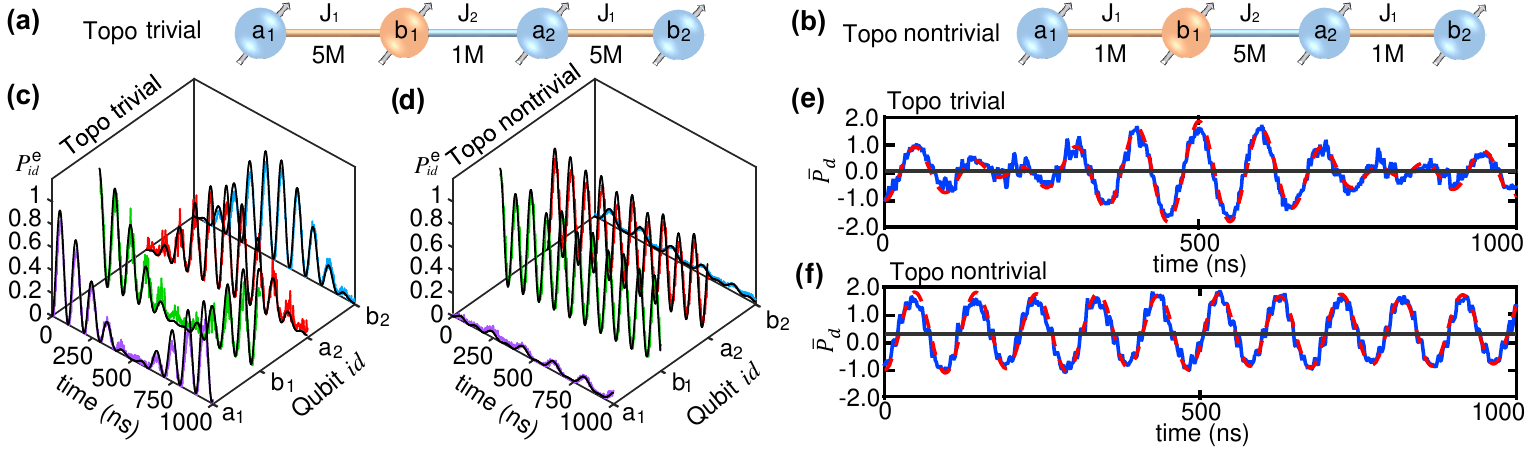}
\caption{\textbf{Topological winding number measurements.} \textbf{a} and \textbf{b} Schematic of the experiments in which only $a_1, b_1, a_2$, and $b_2$ are used without $a_3$.
The couplings between neighboring qubits are configured into $J_1$-$J_2$-$J_1$ = 5-1-5~(MHz) (\textbf{a}, topologically trivial) and $J_1$-$J_2$-$J_1$ = 1-5-1~(MHz) (\textbf{b}, topologically
nontrivial), respectively. \textbf{c} and \textbf{d} Time evolution of the qubit excitation $P_{id}^{\mathrm{e}}$ for $id=a_1, b_1, a_2, b_2$ for the two different coupling configurations.
Dots are experimental data while solid lines are calculated from the ideal Hamiltonian (Eq.~\ref{Eq:DSC}) with the measured system decoherence for an initial state
$\ket{gegg}$. \textbf{e} and \textbf{f} Time evolution of $\bar{P}_d=(P_{a_1}^\mathrm{{e}}-P_{b_1}^\mathrm{{e}})+2(P_{a_2}^\mathrm{{e}}-P_{b_2}^\mathrm{{e}})$ for the two cases. Dots are
experimental data (averaged 5000 times), red dashed lines are from numerical simulations, and the black horizontal lines represent the oscillation centers.}
\label{fig:topo_invariant}
\end{figure*}

We implement the experiment in a superconducting circuit~\cite{You2011,Devoret2013,Gu2017} consisting of five cross-shaped transmon qubits (Xmons, $a_1, b_1, a_2, b_2$, and $a_3$)~\cite{BarendsPRL2013,Barends2014} arranged in a linear array with fixed capacitive nearest-neighbor couplings, as shown in Fig.~\ref{fig:Model}\textbf{b}. Each qubit has independent $XY$ and $Z$ controls. Separate $\lambda/4$ resonators with different frequencies couple to individual qubits for independent readouts. The average qubit $T_1\approx 18~\mu$s and $T^*_2\approx 17~\mu$s at the frequency sweet spots. We use a Josephson parametric amplifier~\cite{Hatridge,Roy2015}, a gain over 20~dB and a bandwidth about 260~MHz, for high-fidelity single-shot measurements of the qubits. To overcome the readout imperfections, we in addition use a calibration matrix to reconstruct the readout results based on Bayes' rule.

The topologically trivial and nontrivial phases require different qubit-qubit coupling configurations, necessitating full control of the effective couplings between neighboring qubits. Tunable couplings through parametrical modulations of the qubit frequencies can be realized in situ without increasing circuit complexity~\cite{Zhou2009,Strand2013,Wu2018,Reagor2018,Li2018}, therefore are ideal for topological simulations. We adopt this technique throughout our experiment to realize the required neighboring qubit coupling strengths as described in Eq.~\ref{Eq:DSC}.

Explicitly, we apply
\begin{eqnarray}
\label{Eq:ParaMod}
\omega_{id}=\omega_{\mathrm{o}\_{id}}+\varepsilon_{id} \sin(\mu_{id}t+\varphi_{id}),
\end{eqnarray}
where $\omega_{\mathrm{o}\_{id}}$ is the mean operating frequency, $\varepsilon_{id}$, $\mu_{id}$, and $\varphi_{id}$ are the modulation amplitude, frequency, and phase, respectively for the qubit $id=a_x, b_x$ in the chain. By neglecting the higher order oscillating terms and under the resonant conditions $\omega_{\mathrm{o}\_{b_x}}-\omega_{\mathrm{o}\_{a_x}}=\mu_{b_x}$ or $\omega_{\mathrm{o}\_{b_{x-1}}}-\omega_{\mathrm{o}\_{a_x}}=\mu_{a_{x}}$, the effective coupling strengths are
\begin{align}
\label{eq:J1J2}
\begin{split}
J_1&=g_{a_x,b_x}\mathcal{J}_1(\alpha_{b_x})\mathcal{J}_0(\alpha_{a_x}) e^{i(\varphi_{b_x}+\pi/2)},
\\
J_2&=g_{b_{x-1},a_x}\mathcal{J}_1(\alpha_{a_x})\mathcal{J}_0(\alpha_{b_{x-1}}) e^{-i(\varphi_{a_x}-\pi/2)},
\end{split}
\end{align}
where $\mathcal{J}_m(\alpha)$ is the $m$th Bessel function of the first kind and $g_{a_x,b_x}$ is the static capacitive coupling strength between neighboring qubits. Both $J_1$ and $J_2$ can be
conveniently tuned via changing $\alpha_{id}=\varepsilon_{id}/\mu_{id}$ of the external modulation. Note that the qubit at the edge (for example, $a_1$) could be stationary without parametric
modulation, while the middle qubit can be parametrically modulated with two independent sinusoidal drives in order to tune the coupling strengths with its two neighboring qubits respectively.
The experimental setup, device parameters, and parametric modulation parameters are all presented in detail in Supplementary Materials.

\begin{figure*}[tb]
\includegraphics{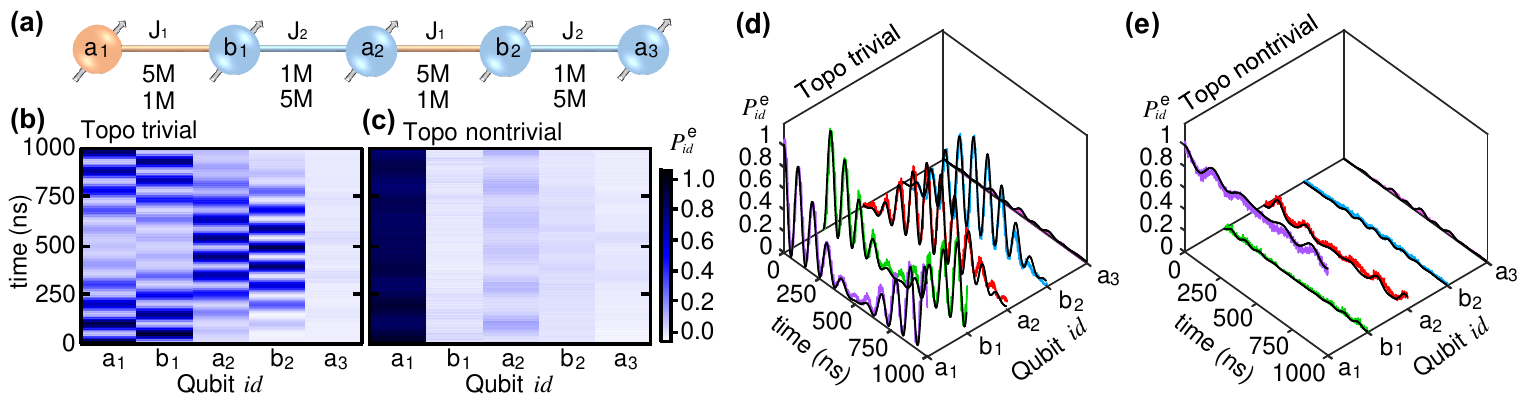}
\caption{\textbf{Observation of topological magnon edge states.} \textbf{a} Schematic of the experiment in which all five qubits have been used. The couplings between neighboring qubits are configured into $J_1$-$J_2$-$J_1$-$J_2$ = 5-1-5-1~(MHz) (topologically trivial) and $J_1$-$J_2$-$J_1$-$J_2$ = 1-5-1-5~(MHz) (topologically nontrivial), respectively. \textbf{b} and \textbf{c} Two-dimensional representation of time evolutions of all qubits' excited state populations [$P_{id}^{\mathrm{e}}(t)$]. \textbf{d} and \textbf{e} Time evolution of $P_{id}^{\mathrm{e}}$. Dots are experimental data while solid lines are calculated from the ideal Hamiltonian (Eq.~\ref{Eq:DSC}) with the measured system decoherence for an initial state $\ket{egggg}$.}
\label{fig:topo_edge}
\end{figure*}

We firstly demonstrate that the topological winding number can be measured by single-magnon quantum dynamics in a chain of four transmon qubits, provided the qubit chain is initially prepared in a single-magnon bulk state. This dynamic method for measuring topological winding number was originally proposed in a linear-optics system for studying discrete-time quantum walk~\cite{Cardano2017}. We choose to excite one of the middle qubits to the excited state $|e\rangle$ and leave the other qubits in the ground state $|g\rangle$, leading to an initial state of the system $|\psi (t=0)\rangle =|gegg\rangle$. After an evolution time $t$, the state of the system becomes $|\psi (t)\rangle =e^{-i\hat{H}t}|\psi (t=0)\rangle$. To reveal the relationship between this dynamics and the topological winding number, we introduce the chiral displacement (CD) operator $\hat{P}_{d}=\sum_{x=1}^{2}x(\hat{P}_{a_{x}}^{e}-\hat{P}_{b_{x}}^{e})$
with $\hat{P}_{id}^{e}=|e\rangle _{id}\langle e|$ ($id=a_{x},b_{x}$). In the long-time limit, the topological winding number $\nu$ can be extracted from the time-averaged CD, $\nu ={\lim_{T\rightarrow \infty }}\frac{2}{T}\int_{0}^{T}dt\,\bar{P}_{d}(t)$, where $T$ is the evolution time and $\bar{P}_{d}(t)=\langle \psi (t)|\hat{P}_{d}|\psi(t)\rangle$ is the CD associated with the dynamics of the single-magnon state (see Supplementary Materials). As we can see, the topological winding number is two times the time-averaged CD, i.e., the oscillation center of the CD versus time. Experimentally, to measure the time-averaged CD we only need to track the time evolution of the excitation for each qubit.

In the experiment, as shown in Figs.~\ref{fig:topo_invariant}\textbf{a} and \ref{fig:topo_invariant}\textbf{b}, we tune the qubit chain into two configurations with the qubit coupling dimerization $J_1>J_2$ and $J_1<J_2$, corresponding to the topologically trivial and nontrivial magnon insulator states, respectively. After preparing the initial state $|\psi (0)\rangle$, we measure the time evolution (with an interval of 1 ns) of the qubit excitation of the four qubits and show the experimental data in Figs.~\ref{fig:topo_invariant}\textbf{c} and \ref{fig:topo_invariant}\textbf{d}. The measured excitation evolutions agree well with the theoretical predictions. Based on these time-resolved excitation data for each qubit, we directly derive the time evolution of CDs and plot them in Figs.~\ref{fig:topo_invariant}\textbf{e} and \ref{fig:topo_invariant}\textbf{f}. Clearly the two curves oscillate around two different center values, qualitatively giving the signature of different topological winding numbers. The evolution time in our experiment is chosen as 1~$\mu$s, during which the experimentally measured time-averaged CDs are $0.015$ and $0.359$ for the topologically trivial and nontrivial cases, respectively. Both experiments agree very well with the theoretically expected values $0$ and $0.378$, giving the experimentally measured topological winding numbers $v=0.030$ and $\nu=0.718$ for the two cases. The measured winding number for the topologically trivial case is quite close to the ideal value.


The reasons for the difference in the topologically nontrivial case between the measured winding number $\nu=0.718$ and the ideal value $\nu=1$ are that both the evolution time and the qubit chain we choose are not long enough and there is also inevitable system decoherence. Nevertheless, our experimental data within 1~$\mu$s agrees excellently with the theoretical expectation and demonstrates the validness of the method using single-magnon dynamics to measure topological winding number. The clear distinction between the measured nontrivial and trivial topological winding numbers thus can unambiguously distinguish the topologically nontrivial and trivial magnon insulator states.


The second hallmark for topological magnon insulator states is the existence of topological magnon edge states at the boundary. When the qubit chain is tuned into the topological magnon insulator state, according to bulk-edge correspondence~\cite{Hasan2010,Qi2011}, topological magnon edge state will emerge at the edges of the qubit chain. The wavefunctions of the left and right magnon edge states can be analytically derived as $|\psi_{L}\rangle=\sum_x(-1)^{x}(J_1/J_2)^{x}\sigma_{a_{x}}^{+}|gg\cdot\cdot gg\rangle$ and $|\psi_{R}\rangle=\sum_x(-1)^{N-x}(J_1/J_2)^{N-x}\sigma_{b_{x}}^{+}|gg\cdot\cdot gg\rangle$, respectively (see Supplementary Materials). It turns out that the magnon in the left (right) edge state only occupies the $a$-type ($b$-type) qubit and is maximally distributed in the leftmost (rightmost) qubit. Such two features provide a mean to observe the topological magnon edge states. However, the coupling between the left and right magnon edge states is very large due to finite lattice size effect, we cannot unambiguously observe the left or right magnon edge state localization in a short qubit chain. This problem can be solved by tuning the qubit chain with an odd number of qubits, where the right topological magnon edge state has been artificially removed (see Supplementary Materials).

Now we show that the left topological magnon edge state can be clearly observed in a chain of five qubits where there is no right topological magnon edge state (see Supplementary Materials). As shown in Fig.~\ref{fig:topo_edge}\textbf{a}, we can tune the qubit couplings in a chain of five qubits to make the system topologically trivial and nontrivial, respectively. Initially, the leftmost $a$-type qubit is excited and a single magnon has been put on the leftmost qubit with the initial system state $|\psi(t=0)\rangle=|egggg\rangle$. Then, we let this magnon state evolve for certain time and measure the time evolution of the magnon density in the qubit chain. The results for the qubit chain being tuned into the topologically trivial state are shown in Fig.~\ref{fig:topo_edge}\textbf{b}. As expected, there is no magnon edge state localization and the wavepacket has a ballistic spread versus time, which is a typical feature of bulk Bloch state. The reason is that the initial magnon state in this case is a superposition of different bulk states, therefore, it evolves in the qubit chain via the bulk state wavepackets and does not support edge state localization.

In contrast, if the qubit chain is tuned into the topologically nontrival state that can support left magnon edge states, as shown in Fig.~\ref{fig:topo_edge}\textbf{c}, the measured magnon density is always maximal in the leftmost qubit. This is because the initial magnon state $|\psi(t=0)\rangle$ has a large overlap with the left magnon edge state $|\psi_L\rangle$. The magnon state thus mainly evolves in the qubit chain based on the edge state wavepacket and always maximally localizes in the leftmost qubit. Moreover, the magnon only populates the $a$-type qubits, also satisfying the feature of left topological magnon edge state as mentioned before. These two features prove the existence of left topological magnon edge state and clearly indicate that the system is topologically nontrivial. In Figs.~\ref{fig:topo_edge}\textbf{d} and \ref{fig:topo_edge}\textbf{e}, we also find that the measured qubit excitation evolutions agree excellently with the theoretical predictions.

\begin{figure}[tb]
\includegraphics{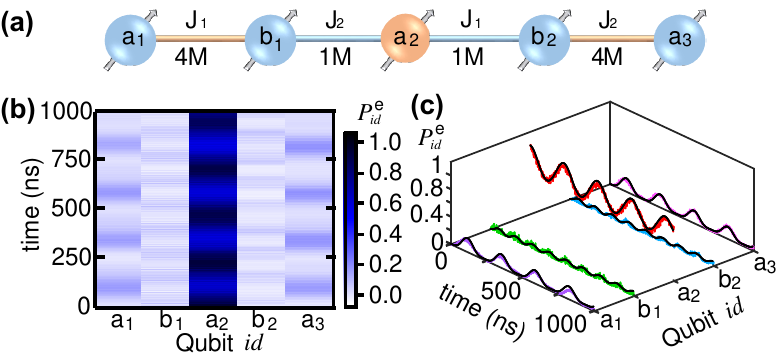}
\caption{\textbf{Observation of topological magnon defect states.} \textbf{a} Schematic of the experiment. The couplings between neighboring qubits are tuned into $J_1$-$J_2$-$J_1$-$J_2$ = 4-1-1-4~(MHz). \textbf{b} Two-dimensional representation of time evolutions of all qubits' excited state populations. \textbf{c} Time evolutions of $P_{id}^{\mathrm{e}}$. Dots are experimental data while solid lines are calculated from the ideal Hamiltonian (Eq.~\ref{Eq:DSC}) with the measured system decoherence for an initial state $\ket{ggegg}$.}
\label{fig:topo_defect}
\end{figure}

The third important topological aspect is the emergence of a topological defect state at the interface between topologically trivial and nontrivial regions~\cite{SSH1979}. When the qubit chain is tuned with two different topological configurations, a topological interface separating the topologically trivial ($J_1>J_2$) and nontrivial ($J_1<J_2$) regions can be created, where a topological magnon defect state is trapped. As shown in Fig.~\ref{fig:topo_defect}\textbf{a}, we can create such a defect state at qubit $a_2$ in a chain of five qubits. The magnon in the topological defect state should only occupy $a$-type qubits and its density should be maximally distributed in qubit $a_2$ (see Supplementary Materials). Experimentally, we initially excite qubit $a_2$ and prepare the system in $|\psi(t=0)\rangle=|ggegg\rangle$. Such an initial state has a large overlap with the wavefunction of the topological magnon defect state. If the system has the topological defect state, the magnon will propagate in the qubit chain via the defect state wavepacket. In the experiment, after evolving $\ket{\psi(t=0)}$ for certain time, we measure the final magnon density distribution in the qubit chain. The experimental results are shown in Fig.~\ref{fig:topo_defect}\textbf{b} and indeed indicate that the magnon is maximally localized in the center qubit $a_2$ and only has populations in the $a$-type qubits, unambiguously demonstrating the existence of a topological magnon defect state. The time evolutions of qubit excitation for the five qubits are also shown in Fig.~\ref{fig:topo_defect}\textbf{c}, agreeing well with the theoretical expectations.


In conclusion, our experiment has demonstrated the potential of a tunable superconducting qubit chain as a versatile platform for exploring topology, including measuring topological invariants and observing topological edge and defect states. Since multiple-qubit excitations can be precisely prepared in this system, our study paves the way for further using a longer qubit chain to realize interacting symmetry-protected topological states~\cite{Chen2011Symmetry,Chen1604} and probe symmetry-protected ground state degeneracy~\cite{Sylvain2018}. Through periodically driving the qubit frequencies, non-equilibrium interacting symmetry-protected topological states also can be studied \cite{PotterPRX2016}. Besides, it is also quite interesting to study how the topological states in the qubit chain helps to accomplish topologically protected quantum information processing tasks~\cite{Mei2018PRA,Nayak2008}. In addition to superconducting qubits, our experiment can be generalized to other qubit systems and could attract broad interests in exploring symmetry-protected topological states with different quantum computing platforms.

\begin{acknowledgments}
\textbf{Acknowledgments.} This work is supported by the National Key Research and Development Program of China (2017YFA0304203, 2017YFA0304303, 2016YFA0301803); Natural National Science Foundation of China (11474177, 11604392, 11434007, 11874156, 61771278); Changjiang Scholars and Innovative Research Team in University of Ministry of Education of China (PCSIRT)(IRT\_17R70); Fund for Shanxi 1331 Project Key Subjects Construction; 111 Project (D18001). LS also thanks R. Vijay and his group for help on the parametric amplifier measurements.
\end{acknowledgments}

\end{document}


\title{Supplementary Materials for ``Observation of topological magnon insulator states in a superconducting circuit"}

\author{W. Cai}
\thanks{These two authors contributed equally to this work.}
\affiliation{Center for Quantum Information, Institute for Interdisciplinary Information Sciences, Tsinghua University, Beijing 100084, China}

\author{J. Han}
\thanks{These two authors contributed equally to this work.}
\affiliation{Center for Quantum Information, Institute for Interdisciplinary Information Sciences, Tsinghua University, Beijing 100084, China}

\author{Feng Mei} \email{meifeng@sxu.edu.cn}
\affiliation{State Key Laboratory of Quantum Optics and Quantum Optics Devices,
    Institute of Laser Spectroscopy, Shanxi University, Taiyuan, Shanxi 030006, China}
\affiliation{Collaborative Innovation Center of Extreme Optics,
Shanxi University, Taiyuan, Shanxi 030006, China}

\author{Y. Xu}
\affiliation{Center for Quantum Information, Institute for Interdisciplinary Information Sciences, Tsinghua University, Beijing 100084, China}

\author{Y. Ma}
\affiliation{Center for Quantum Information, Institute for Interdisciplinary Information Sciences, Tsinghua University, Beijing 100084, China}

\author{X. Li}
\affiliation{Center for Quantum Information, Institute for Interdisciplinary Information Sciences, Tsinghua University, Beijing 100084, China}

\author{H. Wang}
\affiliation{Center for Quantum Information, Institute for Interdisciplinary Information Sciences, Tsinghua University, Beijing 100084, China}

\author{Y. P. Song}
\affiliation{Center for Quantum Information, Institute for Interdisciplinary Information Sciences, Tsinghua University, Beijing 100084, China}

\author{Zheng-Yuan Xue} \affiliation{Guangdong Provincial Key Laboratory of Quantum Engineering
and Quantum Materials, and School of Physics and Telecommunication Engineering,
South China Normal University, Guangzhou 510006, China}

\author{Zhang-qi Yin}
\affiliation{Center for Quantum Information, Institute for Interdisciplinary Information Sciences, Tsinghua University, Beijing 100084, China}

\author{Suotang Jia}
\affiliation{State Key Laboratory of Quantum Optics and Quantum Optics Devices,
    Institute of Laser Spectroscopy, Shanxi University, Taiyuan, Shanxi 030006, China}
\affiliation{Collaborative Innovation Center of Extreme Optics,
Shanxi University, Taiyuan, Shanxi 030006, China}

\author{Luyan Sun}\email{luyansun@tsinghua.edu.cn}
\affiliation{Center for Quantum Information, Institute for Interdisciplinary Information Sciences, Tsinghua University, Beijing 100084, China}

\maketitle

\section{Experimental Device, Setup, and Sequences}

\begin{figure*}[hbt]
{\centering\includegraphics[width=0.90\textwidth]{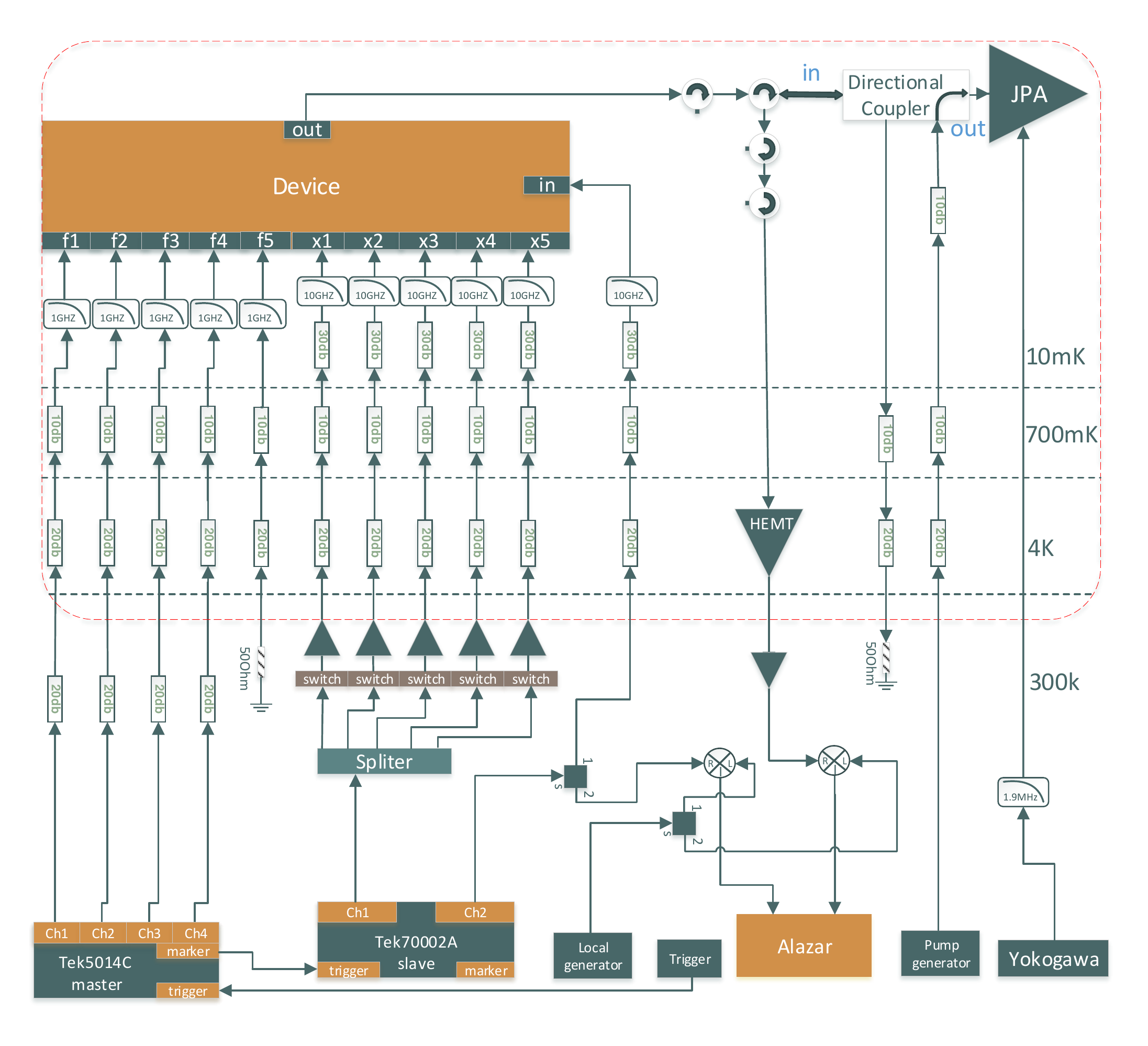}}
\caption{\textbf{Details of wiring and circuit components.} The experimental device consists of five cross-shaped transmon qubits (Xmons, $a_1, b_1, a_2, b_2$, and $a_3$)~\cite{BarendsPRL2013,Barends2014} arranged in a linear array. Each qubit has independent $XY$ and $Z$ controls which are properly attenuated and low-pass filtered. Separate $\lambda/4$ resonators with different frequencies couple to individual qubits for independent and simultaneous readouts. One four-channel AWG5014C is used to fully manipulate the flux biases of the first four qubits ($a_1, b_1, a_2$, and $b_2$) while the flux-bias line of the fifth one $a_3$ is terminated with 50 ohm at its sweet spot. One two-channel AWG70002A, synchronized with AWG5014C, is used to realize all $XY$ controls and readouts of the qubits. Because of its large bandwidth and sampling rate, AWG70002A can directly generate the qubit control and readout pulses without extra IQ modulations. The $XY$ manipulation signal is divided and managed through separate RF switches for selective control of individual qubits. A JPA at 10~mK with a gain over 20~dB and a bandwidth about 260~MHz is used for high-fidelity single-shot measurements of the qubits. A high-electron-mobility-transistor (HEMT) amplifier at 4 K and an amplifier at room temperature are also used before the down-conversion of the readout signal to different frequencies with a different generator as LO. To eliminate the readout signal phase fluctuation, part of the readout signal does not go through the refrigerator and is down-converted as a reference to lock the phase of the returning readout signal from the device.}
\label{fig:experimentsetup}
\end{figure*}

Our experimental chip is anchored in an aluminum sample holder and measured in a dilution refrigerator with a base temperature of about 10~mK. An additional magnetic shield is used to cover the device for a clean electromagnetic environment. Figure~\ref{fig:experimentsetup} shows the measurement circuitry. For full frequency manipulation of the qubits, we use one four-channel arbitrary waveform generators AWG5014C to control the flux-biases of the qubits $a_1, b_1, a_2$, and $b_2$. This allows individual parametric modulation of each qubit frequency. The flux control line of $a_3$ is terminated with 50 ohm at room temperature and its frequency is at the sweet spot.

Due to the ground plane return currents, there are inevitable crosstalks (the maximum one in our device is about 10\%) between flux-bias lines and qubits. This crosstalk can be corrected by orthogonalization of the flux-bias lines through an inversion of the qubit frequency response matrix, leading to independent control of only the desired qubit without changing the other qubit frequencies. Since $a_3$ is biased at its sweet spot and not sensitive to the crosstalk from other qubits' flux control, we do the orthogonalization of the flux-bias lines only for $a_1, b_1, a_2$, and $b_2$, which appears sufficient for our experiment.

To achieve frequency modulations and fast switches between the idle and operating points, it is necessary to change the flux biases in fast time scale. However, the control circuit to generate the control pulses and wiring outside and inside the refrigerator cause finite rise time and ringing of the flux-control pulses, which need to be carefully calibrated out. We use the deconvolution method to correct the unwanted response in the control system based on the measured response function of the control circuit.


A two-channel AWG70002A, synchronized with AWG5014C, is used to realize all $XY$ controls and readouts of the qubits. Because of its large bandwidth and sampling rate, AWG70002A can directly generate the qubit control pulses without extra IQ modulations. In our experiment, the control of the five qubits do not need to be on simultaneously, therefore we use fast switches to manage the individual control of each qubit. The readout signals for individual qubits are also directly generated from AWG70002A without extra IQ modulations.

A Josephson parametric amplifier (JPA)~\cite{Hatridge,Roy2015} at 10~mK is used as the first stage of the transmitted readout signal amplification. The JPA has a gain over 20~dB and a bandwidth about 260~MHz, therefore, it allows for high-fidelity single-shot measurements of all qubits individually and simultaneously. The readout frequencies of the five qubits are designed to span a range of about 80~MHz, well within the bandwidth of the JPA. Mainly due to the mismatch of the dispersive shifts and the readout resonator decay rates, the two Gaussians in each qubit's readout histograms, corresponding to the ground state $\ket{g}$ and the excited state $\ket{e}$, are not perfectly separated. To overcome this readout imperfection, we use a calibration matrix to reconstruct the readout results based on Bayes' rule.

Readout resonator frequencies, qubit frequencies, qubit coherence times, coupling strengths, and readout resonator decay rates are all presented in Table~\ref{Table:parameters}. The device fabrication, the orthogonalization and deconvolution of the flux-bias lines, the readout histograms, and the calibration matrix to reconstruct the readout results based on Bayes' rule are all described in detail in Ref.~\onlinecite{Li2018}.

\begin{figure}[hbt]
\includegraphics{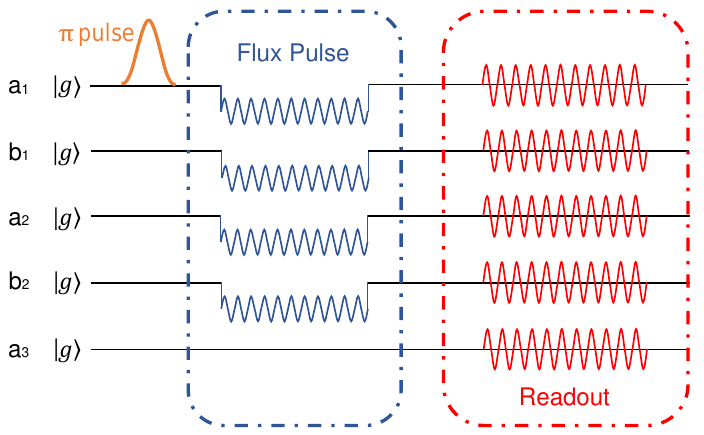}
\caption{\textbf{Experimental sequence for the topologically nontrivial edge state measurement.} This is an example of the experimental sequence for Fig.~3\textbf{c} in the main text. All five qubits are initially at their idle points and $a_1$ is prepared in the excited state by a $\pi$ pulse. In the following, fast step pulses are used to bias the qubits (except for $a_3$) from their idle points to the operating points quickly and then the frequency modulations are on simultaneously to achieve the required coupling configuration for various time $t$. In the end, fast step pulses immediately return all qubits back to their idle points for simultaneous final qubit state measurements to get $P_{id}^{\mathrm{e}}(t)$.}
\label{fig:sequence}
\end{figure}

In our experiment, all qubits are initially at their idle points and only one of the qubit (dependent on the particular experiment) is prepared in the excited state with a $\pi$ pulse. Then fast step pulses are used to bias the qubits from their idle points to the operating points quickly, followed by simultaneous modulations of the necessary qubits to achieve the required coupling configuration for various time $t$. In the end, fast step pulses immediately return all qubits back to their idle points for simultaneous final qubit state measurements to get $P_{id}^{\mathrm{e}}(t)$ ($id=a_1, b_1, a_2, b_2$, and $a_3$). Figure~\ref{fig:sequence} shows an example of the experimental sequence for the topological edge state measurement. Table~\ref{Table:ModulationParameters} shows the parameters of the parametric modulations to realize the required coupling configurations for the specified experiments. Note that for the experiment to measure the topological defect state, qubit $a_2$ is modulated simultaneously with two sets of $\epsilon$ and $\mu$. When we measure the topologically trivial edge states (Fig.~3\textbf{b} of the main text), we use almost the identical coupling configuration (see Table~\ref{Table:ModulationParameters}) as that for the topologically nontrivial edge states (Fig.~3\textbf{c} of the main text), but with the initial excitation on $a_3$ instead of $a_1$, i.e., the labellings of the five qubits are reversed and so is the coupling configuration (seen from the perspective of the initial excitation). This is done only for reasons of simplicity without changing any underlying physics.

\begin{table*}[bht]
\caption{\textbf{Device parameters.}}
\begin{tabular}{cp{0.9cm}<{\centering}p{0.9cm}<{\centering}p{0.9cm}<{\centering}p{0.9cm}<{\centering}p{0.9cm}<{\centering}p{0.9cm}<{\centering}p{0.9cm}<{\centering}p{0.9cm}<{\centering}p{0.9cm}<{\centering}p{0.9cm}<{\centering}p{0.9cm}<{\centering}}
&&&&&&&&&& \tabularnewline
\hline
\hline
Parameters &\multicolumn{2}{c}{$a_1$}  &\multicolumn{2}{c}{$b_1$}  &\multicolumn{2}{c}{$a_2$}  &\multicolumn{2}{c}{$b_2$}  &\multicolumn{2}{c}{$a_3$} \tabularnewline
\hline
Readout frequency (GHz) &\multicolumn{2}{c}{$6.839$} &\multicolumn{2}{c}{$6.864$} &\multicolumn{2}{c}{$6.879$} &\multicolumn{2}{c}{$6.901$} &\multicolumn{2}{c}{$6.919$} \tabularnewline
Qubit frequency (GHz) (sweet spot) &\multicolumn{2}{c}{$4.811$} &\multicolumn{2}{c}{$5.156$} &\multicolumn{2}{c}{$4.901$} &\multicolumn{2}{c}{$5.183$} &\multicolumn{2}{c}{$4.602$} \tabularnewline
$T_1$ ($\mu$s) (sweet spot) &\multicolumn{2}{c}{$20.0$} &\multicolumn{2}{c}{$17.0$} &\multicolumn{2}{c}{$14.8$} &\multicolumn{2}{c}{$17.9$} &\multicolumn{2}{c}{$20.0$} \tabularnewline
Ramsey $T^*_2$ ($\mu$s) (sweet spot) &\multicolumn{2}{c}{$18.5$} &\multicolumn{2}{c}{$16.0$} &\multicolumn{2}{c}{$17.0$} &\multicolumn{2}{c}{$15.0$} &\multicolumn{2}{c}{$19.9$} \tabularnewline
Anharmonicity (MHz) &\multicolumn{2}{c}{$199.70$} &\multicolumn{2}{c}{$181.53$} &\multicolumn{2}{c}{$196.77$} &\multicolumn{2}{{c}}{$212.05$} &\multicolumn{2}{{c}}{$188.13$} \tabularnewline
Adjacent qubit coupling strength ${\rm g}_{j}/2\pi$ (MHz) & &\multicolumn{2}{c}{$16.70$} &\multicolumn{2}{c}{$17.50$} &\multicolumn{2}{c}{$17.50$} &\multicolumn{2}{c}{$16.85$} & \tabularnewline
Qubit-readout dispersive shift $\rm\chi_{qr}/2\pi$ (MHz) &\multicolumn{2}{c}{$0.17$} &\multicolumn{2}{c}{$0.26$} &\multicolumn{2}{c}{$0.20$} &\multicolumn{2}{{c}}{$0.20$} &\multicolumn{2}{{c}}{$0.12$} \tabularnewline
Readout resonator decay rate $\rm\kappa_r/2\pi$ (MHz) &\multicolumn{2}{c}{$0.88$} &\multicolumn{2}{c}{$1.06$} &\multicolumn{2}{c}{$1.23$} &\multicolumn{2}{{c}}{$0.88$} &\multicolumn{2}{{c}}{$0.85$} \tabularnewline
\hline
\hline
\end{tabular} \vspace{-6pt}
\label{Table:parameters}
\end{table*}

\begin{table*}[bht]
\caption{\textbf{Parameters of the parametric modulations to realize the required coupling configurations for various experiments.}}
\begin{tabular}{cp{0.9cm}<{\centering}p{0.9cm}<{\centering}p{0.9cm}<{\centering}p{0.9cm}<{\centering}p{0.9cm}<{\centering}p{0.9cm}<{\centering}p{0.9cm}<{\centering}p{0.9cm}<{\centering}p{0.9cm}<{\centering}p{0.9cm}<{\centering}p{0.9cm}<{\centering}}
&&&&&&&&&& \tabularnewline
\hline
\hline
Parameters &\multicolumn{2}{c}{$a_1$}  &\multicolumn{2}{c}{$b_1$}  &\multicolumn{2}{c}{$a_2$}  &\multicolumn{2}{c}{$b_2$}  &\multicolumn{2}{c}{$a_3$} \tabularnewline
\hline
Qubit operating frequency (GHz) &\multicolumn{2}{c}{$4.811$} &\multicolumn{2}{c}{$5.120$} &\multicolumn{2}{c}{$4.901$} &\multicolumn{2}{c}{$4.680$} &\multicolumn{2}{c}{$4.602$} \tabularnewline
Qubit center frequency for parametric modulation (GHz) &\multicolumn{2}{c}{$4.760$} &\multicolumn{2}{c}{$4.940$} &\multicolumn{2}{c}{$4.830$} &\multicolumn{2}{c}{$4.680$} &\multicolumn{2}{c}{$4.602$} \tabularnewline
$\varepsilon$ for topo. defect state (MHz) && &\multicolumn{2}{c}{$131.52$} &\multicolumn{2}{c}{$14.73,38.38$} &\multicolumn{2}{c}{$38.17$} & \tabularnewline
$\mu$ for topo. defect state (MHz) & & &\multicolumn{2}{c}{$181.19$} &\multicolumn{2}{c}{$107.16,162.23$} &\multicolumn{2}{c}{$70.71$} \tabularnewline
$\varepsilon$ for topo. trivial edge state (MHz) &\multicolumn{2}{c}{$34.82$} &\multicolumn{2}{c}{$77.33$} &\multicolumn{2}{{c}}{$40.03$} &\multicolumn{2}{{c}}{$49.17$} \tabularnewline
$\mu$ for topo. trivial edge state (MHz) &\multicolumn{2}{c}{$171.04$} &\multicolumn{2}{c}{$100.62$} &\multicolumn{2}{{c}}{$161.66$} &\multicolumn{2}{{c}}{$71.10$} \tabularnewline
$\varepsilon$ for topo. nontrivial edge state (MHz) &\multicolumn{2}{c}{$35.02$} &\multicolumn{2}{c}{$76.83$} &\multicolumn{2}{{c}}{$40.48$} &\multicolumn{2}{{c}}{$49.17$} \tabularnewline
$\mu$ for topo. nontrivial edge state (MHz) &\multicolumn{2}{c}{$171.38$} &\multicolumn{2}{c}{$100.92$} &\multicolumn{2}{{c}}{$161.71$} &\multicolumn{2}{{c}}{$71.75$} \tabularnewline
$\varepsilon$ for topo. nontrivial winding number (MHz) &\multicolumn{2}{c}{$37.03$} &\multicolumn{2}{c}{$77.02$} &\multicolumn{2}{{c}}{$36.30$} \tabularnewline
$\mu$ for topo. nontrivial winding number (MHz) &\multicolumn{2}{c}{$171.59$} &\multicolumn{2}{c}{$102.57$} &\multicolumn{2}{{c}}{$160.92$}
\tabularnewline
\hline
Operating frequency for topo. trivial winding number (GHz) &\multicolumn{2}{c}{$4.811$} &\multicolumn{2}{c}{$5.128$} &\multicolumn{2}{c}{$4.901$} &\multicolumn{2}{c}{$5.183$}  \tabularnewline
Center frequency for topo. trivial winding number (GHz) &\multicolumn{2}{c}{$4.780$} &\multicolumn{2}{c}{$4.930$} &\multicolumn{2}{c}{$4.830$} &\multicolumn{2}{c}{$4.990$} \tabularnewline
$\varepsilon$ for topo. trivial winding number (MHz)&& &\multicolumn{2}{c}{$140.65$} &\multicolumn{2}{c}{$17.72$} &\multicolumn{2}{{c}}{$129.12$} \tabularnewline
$\mu$ for topo. trivial winding number (MHz)& &&\multicolumn{2}{c}{$156.18$} &\multicolumn{2}{c}{$108.74$} &\multicolumn{2}{{c}}{$172.71$}\tabularnewline
\hline
\hline
\end{tabular} \vspace{-6pt}
\label{Table:ModulationParameters}
\end{table*}

\section{Topological magnon insulator states in a qubit chain}
In the experiment, we realize a dimerized spin chain model in a superconducting qubit chain, where each unit cell contains two qubits labelled by $a$ and $b$. The resulted qubit chain can be described by the dimerized spin chain Hamiltonian, Eq.~1 in the main text. We omit the constant qubit frequencies and only consider a singe-qubit excitation. Because the number of excitations is conserved in our model, the dimerized spin chain Hamiltonian can be reduced to the single-excitation subspace. In condensed matter physics, this single excitation is called a single magnon~\cite{FukuharaNature2013,Fukuhara2013}. Based on the Matsubara-Matsuda transformation~\cite{Matsubara1956}, the qubit chain can be described with the following magnon Hamiltonian
\begin{equation}
\hat{H_1}=\sum_{x=1}^{N}(J_{1}\hat{a}_{x}^{\dag }\hat{b}_{x}+J_{2}\hat{b}%
_{x}^{\dag }\hat{a}_{x+1}+\text{H.c.),}  \label{DSC}
\end{equation}%
where $\hat{\alpha}_{x}^{\dag }=\hat{\sigma}_{\alpha_{x}}^{+}$ ($\alpha=a,b$) is the magnon creation operator for qubit at $a_{x}$ ($b_{x}$), $J_1$ and $J_2$ are the intracell and intercell qubit couplings, respectively, and $N$ is the number of unit cells. To study its topological feature, we rewrite Eq.~\ref{DSC} in the momentum space as
$\hat{H_1}=\sum_{k_{x}}\hat{m}_{k_{x}}^{\dag }\hat{h}(k_{x})\hat{m}_{k_{x}}$, where $\hat{m}_{k_{x}}=(\hat{a}_{k_{x}},\hat{b}_{k_{x}})^{T}$, $\hat{a}_{k_{x}}$ and $\hat{b}_{k_{x}}$ are the momentum space operators, and
\begin{equation}
\hat{h}(k_{x})=d_{x}\hat{\tau}_{x}+d_{y}\hat{\tau}_{y},
\label{hkDSC}
\end{equation}
with $d_{x}=J_{1}+J_{2}\cos (k_{x})$, $d_{y}=J_{2}\sin (k_{x})$, and $\hat{\tau}_{x}$ and $\hat{\tau}_{y}$ being the Pauli spin operators defined in the momentum space. The topology of the dimerized qubit chain in the single excitation case is same as of the Su-Schrieffer-Heeger model~\cite{SSH1979} and can be characterized by the topological winding number
\begin{equation}
\nu =\frac{1}{2\pi }\int dk_{x}\mathbf{n}\times \partial _{k_{x}} ,
\label{wn}
\end{equation}%
where $\mathbf{n}=(n_{x},n_{y})=(d_{x},d_{y})/\sqrt{d_{x}^{2}+d_{y}^{2}}$. Through a straightforward calculation, we find that
\begin{equation}
\nu =%
\begin{cases}
1, & \mbox{$J_1<J_2$;} \\
0\text{,} & \mbox{$J_1>J_2$ }.%
\end{cases}%
\end{equation}%
The winding number $\nu=1$ ($0$) shows that the above qubit chain is in the topologically nontrivial (trivial) magnon insulator state.

\section{The relationship between the single-magnon dynamics and the topological winding number}

It has been previously demonstrated that topological winding number can be dynamically detected via single-particle discrete- and continuous-time quantum dynamics~\cite{Cardano2017,Maffei2018}. In the following, we will show that such a method also can be used in a chain of superconducting qubits~\cite{Mei2018}. We choose to excite one of the middle qubits to the excited state $|e\rangle$ and the other qubits are all in the ground state $|g\rangle$. Thus, the initial state of the system is
\begin{equation}
|\psi (0)\rangle =|gg\cdots e \cdots gg\rangle.
\end{equation}
The quantum dynamics of this single-excitation state is governed by the Hamiltonian in Eq.~\ref{DSC}. After an evolution time $t$, the state of the system becomes
\begin{equation}
|\psi (t)\rangle =e^{-i\hat{H_1}t}|\psi (0)\rangle.
\end{equation}
The relation between the above quantum dynamics and the
topological feature of the qubit chain can be revealed
through the chiral displacement (CD) in the qubit chain. The CD operator is defined as
\begin{equation}
\hat{P}_{d}=\sum_{x=1}^{N}x(\hat{P}_{a_{x}}^{e}-\hat{P}_{b_{x}}^{e}),
\end{equation}%
with $\hat{P}_{id}^{e}=|e\rangle _{id}\langle e|$ ($id=a_{x},b_{x}$). Then the
time-dependent average of the CD associated with the above single-excitation quantum dynamics is given by
\begin{equation}
\bar{P}_{d}(t)=\langle \psi (t)|\hat{P}_{d}|\psi(t)\rangle.
\label{pdt}
\end{equation}
Furthermore, we transfer Eq.~\ref{pdt} into the momentum space and get
\begin{equation}
\bar{P}_{d}(t)=\frac{1}{2\pi }\int_{-\pi }^{\pi }dk_{x}\langle \psi (0)|e^{i%
\hat{h}(k_{x})t}i\partial
_{k_{x}}\hat{\tau}_{z}e^{-i\hat{h}(k_{x})t}|\psi (0)\rangle.
\label{pdtks}
\end{equation}%
By substituting Eq.~\ref{hkDSC} into Eq.~\ref{pdtks}, we find
$\bar{P}_{d}(t)$ is tied closely with the topological winding
number $\nu$ defined in Eq.~\ref{wn}, i.e.,
\begin{equation}
\begin{split}
\bar{P}_{d}(t)&=\frac{\nu }{2}-\frac{1}{4\pi}\int dk_{x}\cos(2d_{t}t)\mathbf{n}\times \partial _{k_{x}}\mathbf{n},
\label{pdk}
\end{split}
\end{equation}
where $d_{t}=\sqrt{J_1^2+J_2^2+2J_1J_2\cos(k_x)}$. In the long time limit, we can obtain the relationship between the winding number and the time-averaged CD~\cite{Cardano2017,Maffei2018,Mei2018}, i.e.,
\begin{equation}
\nu ={\lim_{T\rightarrow \infty }}\frac{2}{T}\int_{0}^{T}dt\,\bar{P}_{d}(t).
\label{pdwn}
\end{equation}%
Therefore, the time-averaged CD value depends on the topology of the band structure of the qubit chain.

\section{Wavefunction of zero-energy topological edge states}

According to the bulk-edge correspondence for topological states, the existence of edge states is a seminal feature associated with topological insulator states. In the following, we will show that the topological qubit chain supports zero-energy topological edge states. Firstly, we show that the wavefunction of the zero-energy state in an dimerized spin chain can be exactly derived, even in the absence of translational invariance. For this purpose, we consider a generalized spin chain model with its Hamiltonian written as
\begin{equation}
\hat{H'}=\sum_{x=1}^{N}(u_x\hat{\sigma}_{a_x}^{\dag }\hat{\sigma}_{b_x}+w_x\hat{\sigma}%
_{b_x}^{\dag }\hat{\sigma}_{a_{x+1}}+\text{H.c.),}
\end{equation}
which breaks the translational invariance. Suppose the wavefunction of the zero-energy state is
\begin{equation}
|\psi_E\rangle=\sum_x [\lambda(a_x)\hat{\sigma}_{a_x}^{\dag }+\lambda(b_x)\hat{\sigma}_{b_x}^{\dag }]|G\rangle,
\label{ewgDSC}
\end{equation}
where $|G\rangle=|gg\cdots \cdots gg\rangle$ is the vacuum magnon state. By substituting this wavefunction into the Schr\"{o}dinger equation $H'|\psi_E\rangle=0$, when $u_x<w_x$ and in the thermodynamic limit, we can get two solutions
\begin{eqnarray}
\lambda(a_x)=\prod_{j=1}^{x-1}-\frac{u_x}{w_x}\lambda(a_1), \lambda(b_x)=0; \nonumber \\
\lambda(b_x)=-\frac{u_N}{w_x}\prod_{j=x+1}^{N-1}-\frac{u_x}{w_x}\lambda(b_N), \lambda(a_x)=0.
\label{ewgDSCs}
\end{eqnarray}

For the standard translational invariant dimerized spin chain model, $u_x=J_1$ and $w_x=J_2$, and we can derive the wavefunctions of the left and right zero-energy edge states as
\begin{eqnarray}
|\psi_L\rangle=\sum_x(-\frac{J_1}{J_2})^x\hat{\sigma}_{a_x}^{\dag}|G\rangle; \nonumber \\
|\psi_R\rangle=\sum_x(-\frac{J_1}{J_2})^{N-x}\hat{\sigma}_{b_x}^{\dag}|G\rangle.
\label{eW}
\end{eqnarray}
It is found that the magnon in the left (right) edge state only occupies the $a$-type ($b$-type) qubit and its density is maximally distributed in the leftmost (rightmost) qubit. This feature has been clearly demonstrated in Fig. 3\textbf{c} of the main text.

\begin{figure}[hbt]
\includegraphics[width=8cm,height=10cm]{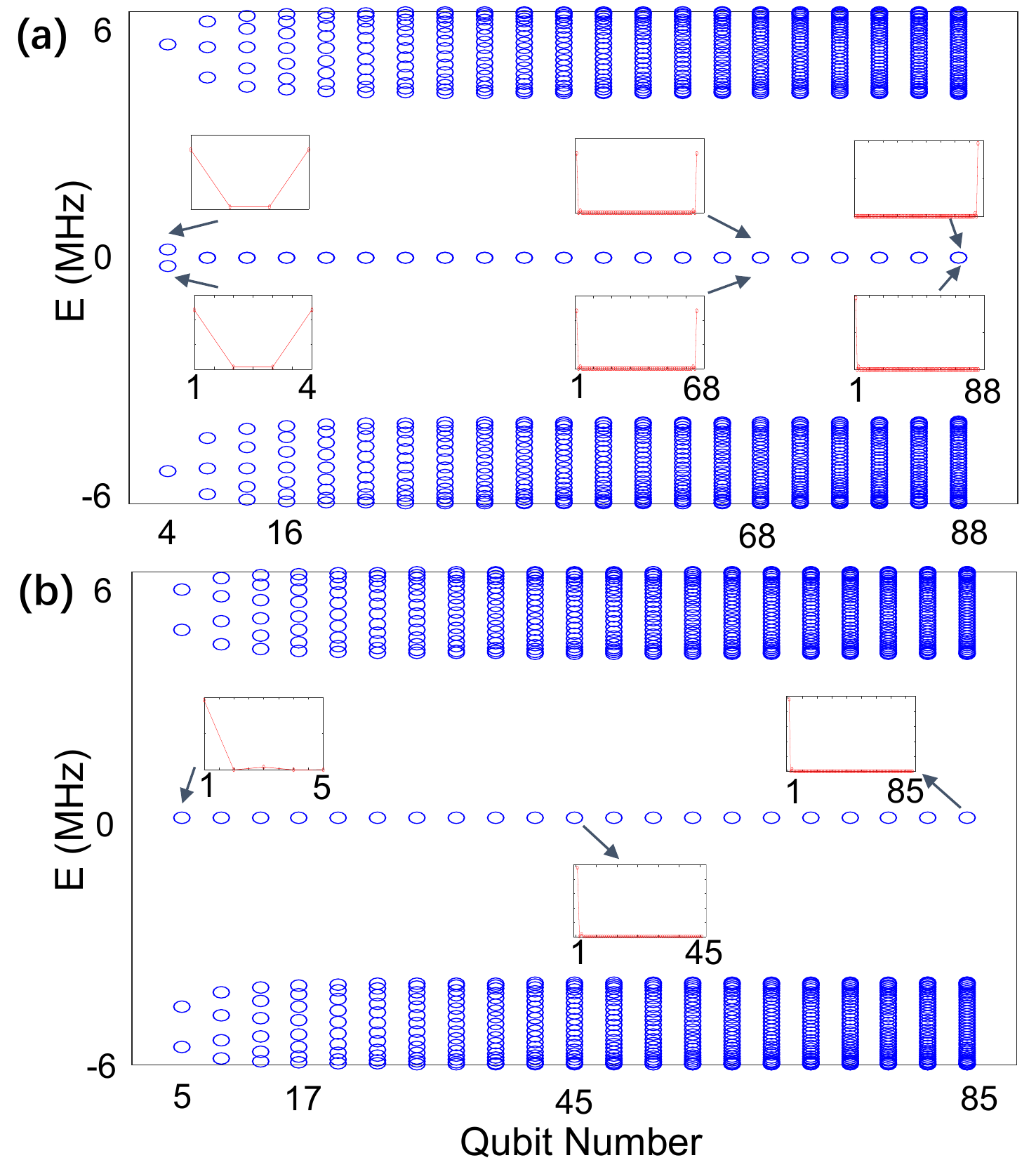}
\caption{The energy spectrum of a qubit chain for even (\textbf{a}) and odd (\textbf{b}) number of qubits. There are two in-gap edge modes for even number of qubits. In contrast, there is only one in-gap edge mode for odd number of qubits. The insets show the density distribution of the edge states. The parameters are chosen as $J_1=1$ MHz and $J_2=5$ MHz. }
\label{fig:QubitSize}
\end{figure}

\section{The influence of qubit lattice size on observing the topological magnon edge states}

\begin{figure}[hbt]
\includegraphics{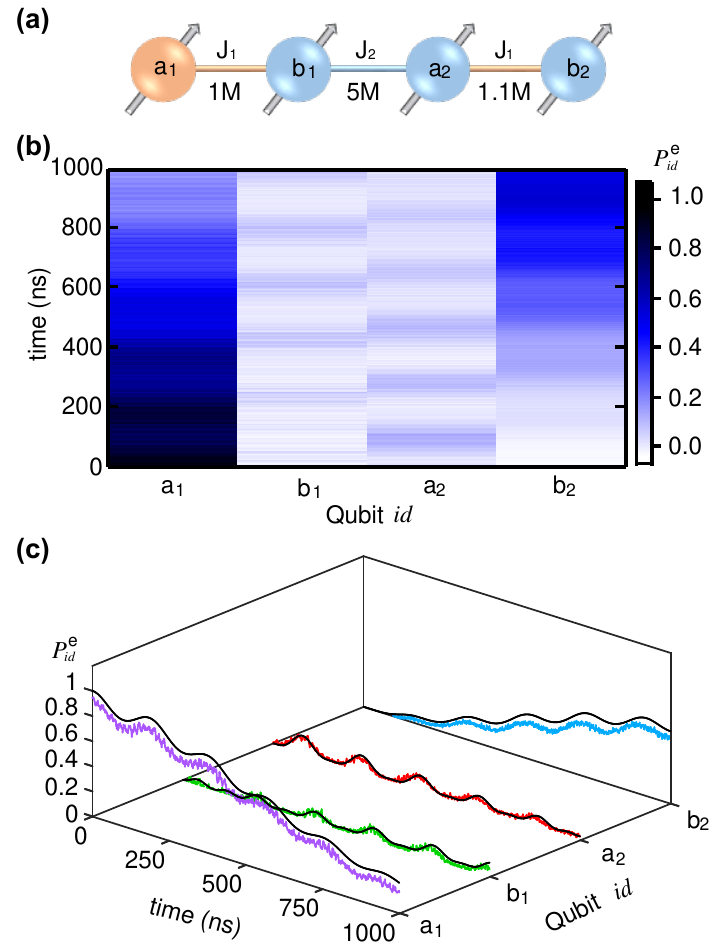}
\caption{\textbf{a} Schematic of the experiment for the edge state measurement in which only four qubits have been used with the excitation initially on $a_1$. The couplings between neighboring qubits are first configured into $J_1$-$J_2$-$J_1$-$J_2$ = 1-5-1-5~(MHz) (topologically nontrivial) as in Fig.~3 of the main text. Then the modulation on the fourth qubit ($b_2$) is turned off, without which $a_3$ is decoupled from the first four qubits and the coupling between $a_2$ and $b_2$ is expected to increase. \textbf{b} Two-dimensional representation of the time evolutions of all four qubits' excited state populations $P_{id}^{\mathrm{e}}(t)$ ($id=a_1, b_1, a_2$, and $b_2$). In this case, the population is not localized on $a_1$ only and can be transferred to $b_2$ as expected, in sharp contrast to the edge state behavior in Fig.~3\textbf{c} of the main text. \textbf{c} Time evolution of $P_{id}^{\mathrm{e}}$. Dots are experimental data while solid lines are calculated from the ideal Hamiltonian (Eq.~1 of the main text) for an initial state $\ket{eggg}$ with the measured system decoherence and the coupling configuration $J_1$-$J_2$-$J_1$ = 1-5-1.1~(MHz).}
\label{fig:4qubits_edgestate}
\end{figure}

In Fig.~\ref{fig:QubitSize}\textbf{a}, we have calculated the energy spectrum of the qubit chain versus even number of qubits.
There are two in-gap topological magnon edge modes in this case. When the qubit chain is very long, we find that the two in-gap edge modes are degenerate with zero energy. The densities of such two modes mostly populate the leftmost and rightmost qubits, yielding the left and right topological magnon edge states $|\psi_L\rangle$ and $|\psi_R\rangle$ respectively. However, due to finite lattice size effect, there is a coupling between the left and the right edge states in a short qubit chain. Suppose the coupling strength is $t_e$, and based on Eq.~\ref{eW} we can get
\begin{equation}
t_e=\langle \psi_L| H|\psi_R\rangle.
\label{edgecoupling}
\end{equation}
This coupling leads to a splitting of the two degenerate zero-energy edge modes into two non-degenerate edge modes. This can be directly seen in Fig.~\ref{fig:QubitSize}\textbf{a}. The result for the four-qubit case shows that there is an energy gap ($\simeq2t_e$) between the two in-gap edge modes. Since the edge state coupling $t_e$ decreases exponentially with the increase of the qubit number, the gap becomes much smaller when the qubit number increases and quickly can not be distinguished, as shown in Fig.~\ref{fig:QubitSize}\textbf{a}. In this case, the wave functions of the two in-gap edge modes are no longer $|\psi_L\rangle$ and $|\psi_R\rangle$, but become hybridizations of these two modes, i.e.
\begin{equation}
|\tilde{\psi}_{L,R}\rangle=(|\psi_L\rangle\pm|\psi_R\rangle)/\sqrt{2}.
\end{equation}
The corresponding densities of the edge states $|\tilde{\psi}_{L}\rangle$ and $|\tilde{\psi}_{R}\rangle$ would not maximally populate the left and right ends of the qubit chain respectively. Instead they both maximally populate the two ends of the qubit chain, as shown in Fig.~\ref{fig:QubitSize}\textbf{a}. For the parameters chosen in Fig.~\ref{fig:4qubits_edgestate}, we numerically find that the critical qubit number when the edge state coupling and the hybridization phenomenon vanish is 86.

In the experiment, we have observed the edge state coupling and the hybridization phenomenon in a qubit chain with four qubits. Fig.~\ref{fig:4qubits_edgestate}\textbf{a} shows the schematic of the experiment. The couplings between neighboring qubits are first configured into $J_1$-$J_2$-$J_1$-$J_2$ = 1-5-1-5~(MHz) (topologically nontrivial) as in Fig.~3 of the main text. Then the modulation on the fourth qubit ($b_2$) is turned off. Due to the vanish of this modulation, $a_3$ is then decoupled from the first four qubits and the coupling between $a_2$ and $b_2$ is also expected to become larger. However, due to the topological protection, this small imperfection does not affect the appearances of the topological magnon edge states.

Initially, we prepare a single-magnon state by exciting the leftmost qubit $a_1$ into an excited state, i.e. $\ket{\psi(t=0)}=|eggg\rangle$. After that, we measure the time evolution of this single-magnon state in the qubit chain. The experimental result is shown in Fig.~\ref{fig:4qubits_edgestate}, which indicates that the magnon firstly mainly populates the leftmost qubit and finally evolves to the rightmost qubit. This is because there is a coupling between the left and right edge states in a short qubit chain, in good agreement with theoretical calculations from the ideal Hamiltonian (Eq.~1 in the main text) for an initial state $\ket{eggg}$ with the coupling configuration $J_1$-$J_2$-$J_1$ = 1-5-1.1~(MHz) and the system decoherence. Thus the coupling between the left and right edge states can not be ignored in a short qubit chain and can lead to edge state hybridization.

A direct method to eliminate the coupling between the left and right edge states and to only observe edge state localization is to perform the experiment in a long qubit chain, which however is not the case for our current device. We solve this problem through an alternative way by using a qubit chain with an odd number of qubits. In this case, the rightmost qubit has been removed, and the system only supports the left edge state while the right edge state has been eliminated. We have numerically demonstrated this point in Fig.~\ref{fig:QubitSize}\textbf{b}. One can find that there is only one in-gap zero-energy mode in a qubit chain with an odd number of qubits, with its density maximally populating the leftmost qubit, regardless of the exact qubit number. In a chain of five qubits, we have experimentally observed that the left magnon edge state only maximally localizes at the leftmost qubit. The experimental results are shown in Fig.~3 of the main text.

\section{Wavefunction of zero-energy topological magnon defect states}

The topological qubit chain also supports the zero-energy topological magnon defect state~\cite{SSH1979}. Such a state is generated at the interface between the topologically trivial ($J_1>J_2$) and nontrivial ($J_1<J_2$) regions. In this case, the qubit lattice breaks the translational invariance. Suppose the interface is located at qubit $a_{x_e}$ and based on Eqs.~\ref{ewgDSC} and \ref{ewgDSCs}, we can derive the wavefunction of the zero-energy topological magnon defect state as
\begin{equation}
|\psi_S\rangle=\Big[\sum_{x< x_e}(-\frac{J_1}{J_2})^{x_e-x}+\sum_{x\geq x_e}(-\frac{J_1}{J_2})^{x-x_e}\Big]\hat{\sigma}_{a_x}^{\dag}|G\rangle. \\
\end{equation}
This shows that the magnon in the topological defect state only occupies $a$-type qubit and its density is maximally distributed in qubit $a_{x_e}$ at the interface. Our experimental data, plotted in Fig.~4 of the main text, show exactly this expected behavior.

\section{Comparison between experimental data and theoretical expectations in two-dimensional plots}

\begin{figure*}[hbt]
\includegraphics[scale=0.95]{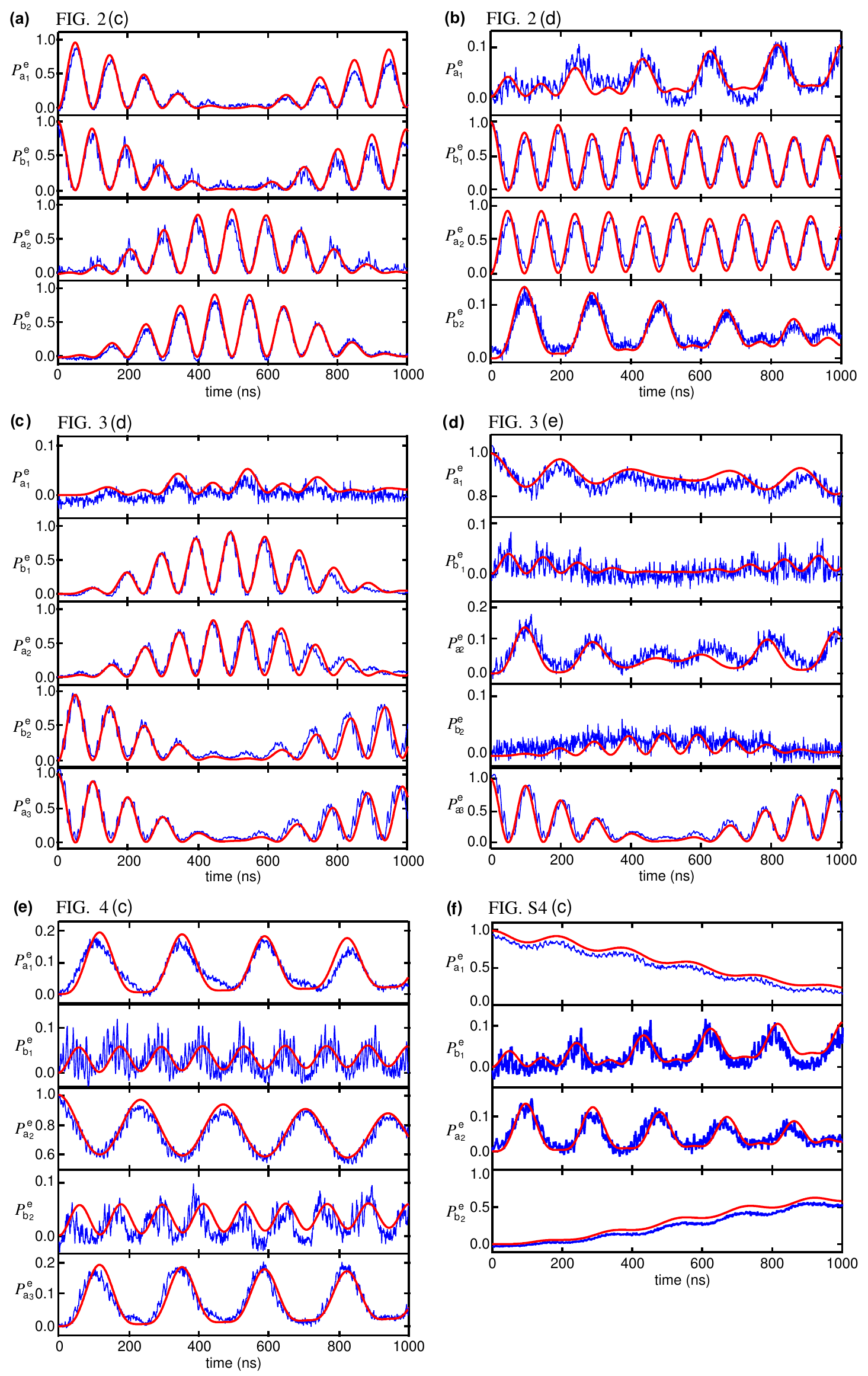}
\caption{\textbf{a-f} correspond to data in Figs.~2(\textbf{c}), 2(\textbf{d}), 3(\textbf{d}), 3(\textbf{e}), and 4(\textbf{c}) of the main text and in Fig.~\ref{fig:4qubits_edgestate}(\textbf{c}), respectively. Dots are experimental data while solid lines are calculated from the ideal Hamiltonian (Eq.~1 in the main text) with measured system decoherence. All experimental data agree well with theoretical expectations. Note that the $y$-axes have different scales.}
\label{fig:2Dplots}
\end{figure*}

To have a clearer comparison between experimental data and theoretical expectations, we show two-dimensional plots in Fig.~\ref{fig:2Dplots} of all data in Figs.~2(\textbf{c}), 2(\textbf{d}), 3(\textbf{d}), 3(\textbf{e}), and 4(\textbf{c}) of the main text and in Fig.~\ref{fig:4qubits_edgestate}(\textbf{c}). All experimental data agree well with theoretical expectations.

\section{More simulations with different coupling configurations}

\begin{figure*}[hbt]
\includegraphics{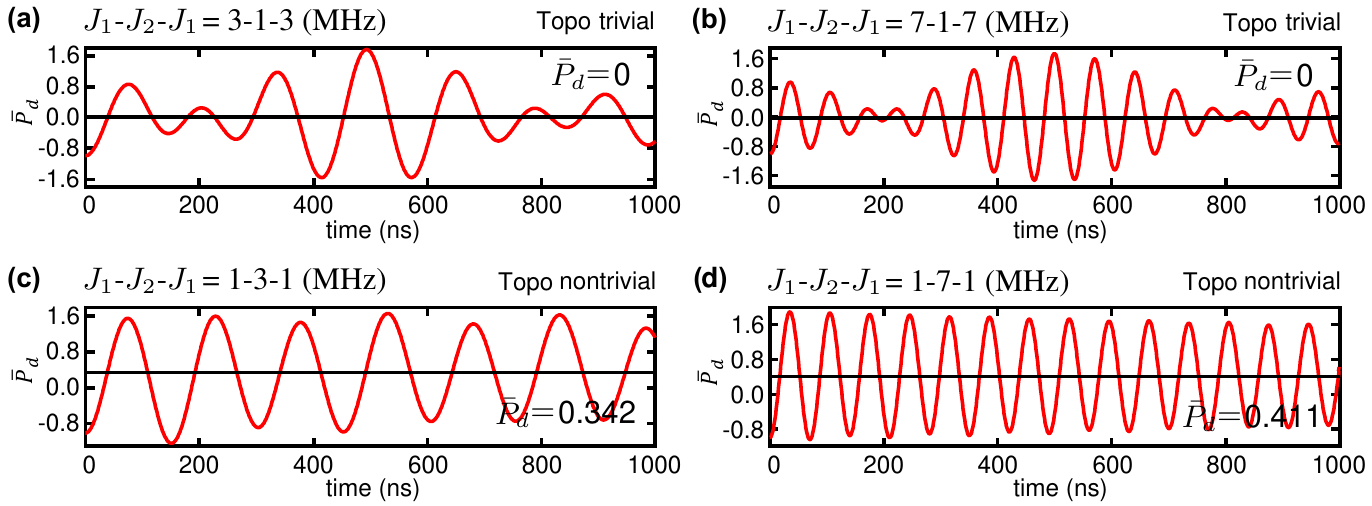}
\caption{Simulated topological winding number measurements with system decoherence for coupling configurations $J_1$-$J_2$-$J_1$=3-1-3 (1-3-1)~(MHz) [\textbf{a} (\textbf{c})] and $J_1$-$J_2$-$J_1$=7-1-7 (1-7-1)~(MHz) [\textbf{b} (\textbf{d})], respectively.}
\label{fig:S_Fig2}
\end{figure*}

\begin{figure*}[hbt]
\includegraphics{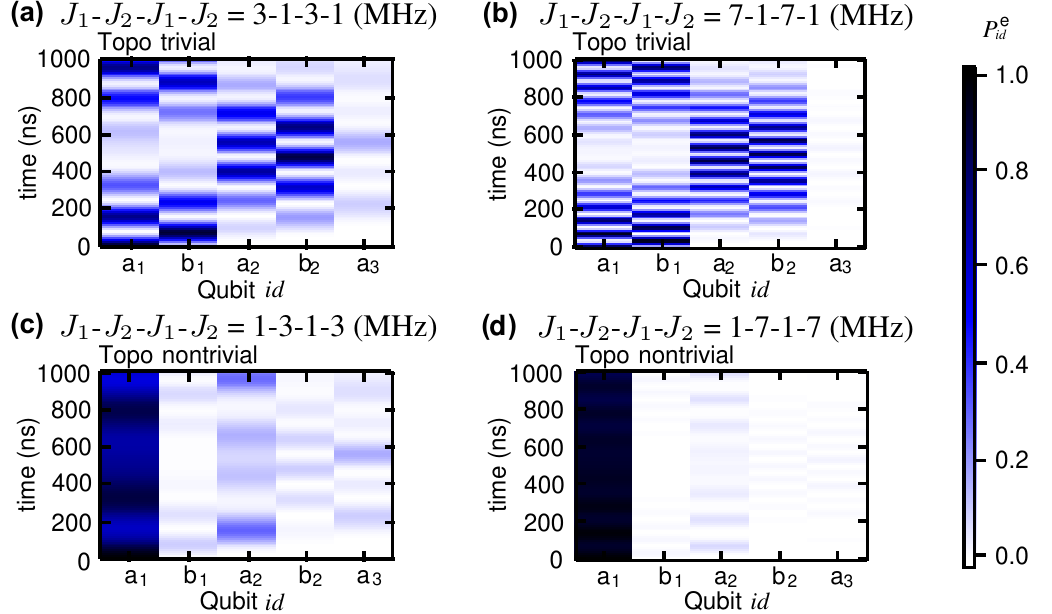}
\caption{Simulated observation of topological magnon edge states with system decoherence for coupling configurations $J_1$-$J_2$-$J_1$-$J_2$=3-1-3-1 (1-3-1-3)~(MHz) [\textbf{a} (\textbf{c})] and $J_1$-$J_2$-$J_1$-$J_2$=7-1-7-1 (1-7-1-7)~(MHz) [\textbf{b} (\textbf{d})], respectively.}
\label{fig:S_Fig3}
\end{figure*}

\begin{figure*}[hbt]
\includegraphics{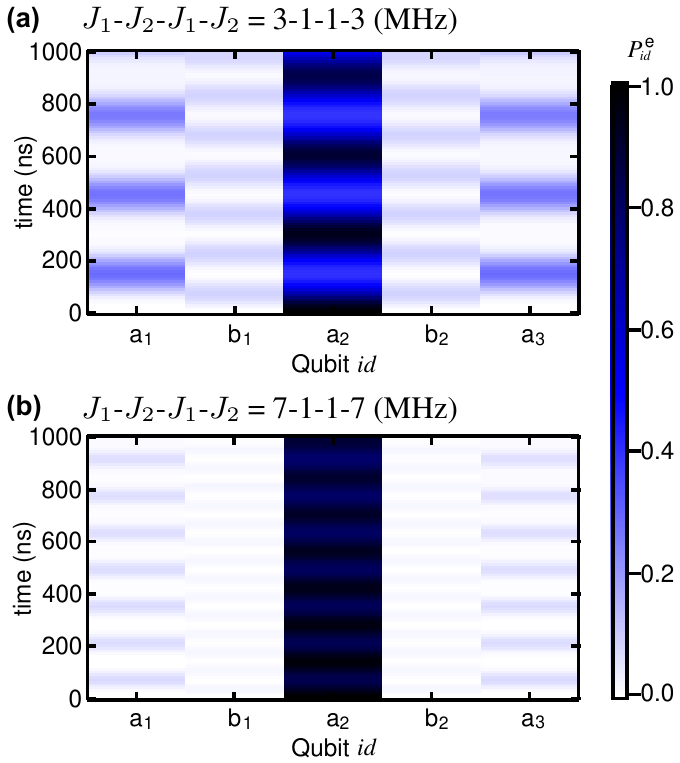}
\caption{Simulated observation of topological magnon defect measurements with system decoherence for coupling configurations $J_1$-$J_2$-$J_1$-$J_2$=3-1-1-3~(MHz) (\textbf{a}) and $J_1$-$J_2$-$J_1$-$J_2$=7-1-1-7~(MHz) (\textbf{b}), respectively.}
\label{fig:S_Fig4}
\end{figure*}

To further illustrate the topological phases shown in the main text, we have performed more detailed simulations with system decoherence for different coupling configurations. The results are shown in Figs.~\ref{fig:S_Fig2}-\ref{fig:S_Fig4} and demonstrate clearly the expected topologically trivial and nontrivial behaviors respectively.

%